\documentclass[iop,apj,appendixfloats,revtex4]{emulateapj}
 
\usepackage[backref,breaklinks,colorlinks,citecolor=blue]{hyperref}
\usepackage[all]{hypcap} %Links go to figures;breaks on deluxetables       

\usepackage{color}

\newcommand{\hst}{\textit{HST}}

\newcommand{\cluster}{SDSS~J1110+6459}
\newcommand{\arcnamelong}{SGAS~J111020.0$+$645950.8}
\newcommand{\arcname}{SGAS~1110}
\newcommand{\zA}{2.481}

\newcommand{\totalmagnification}{$28\pm8$}
\newcommand{\smallestscale}{$r=30$~pc}  % Rounded from 29.
\newcommand{\largestclump}{$r=50$~pc}   % Rounded from 49
\newcommand{\rangeofscales}{$r=30$--50~pc}  % ditto
\newcommand{\fluxcompleteness}{$m_{AB}=33.2$ in F606W}
\newcommand{\HaParatio}{8.45}  % JRR:  $8.45$ from Tab 4.4 of Ferland \& Osterbrock, for T=10^4, ne=100

\newcommand{\cc}{\hbox{cm$^{-3}$}}
\newcommand{\kms}{\hbox{km~s$^{-1}$}}
\newcommand{\Msol}{M$_{\odot}$}
\shorttitle{}
\slugcomment{Accepted for publication in The Astrophysical Journal Letters}
\shortauthors{Johnson et al.}

\begin{document}

%==================================================================================
%   TITLE, AUTHORS, AFFILIATIONS
%==================================================================================
\title{Star Formation at $z=2.481$ in the Lensed Galaxy SDSS J1110$+$6459:\\
Star Formation down to 30~parsec scales}

\author{Traci L. Johnson\altaffilmark{1}, Jane R.~Rigby\altaffilmark{2}, Keren Sharon\altaffilmark{1}, 
Michael D.~Gladders\altaffilmark{3,4}, Michael Florian\altaffilmark{2}, Matthew B.~Bayliss\altaffilmark{5}, 
Eva Wuyts\altaffilmark{6}, Katherine E.~Whitaker\altaffilmark{7,8}$^\dagger$, 
Rachael Livermore\altaffilmark{9}, \& Katherine T. Murray\altaffilmark{10}}

\email{tljohn@umich.edu}
\altaffiltext{*}{Based on observations made with the NASA/ESA Hubble Space Telescope, obtained at the Space Telescope Science Institute, which is operated by the Association of Universities for Research in Astronomy, Inc., under NASA contract NAS 5-26555. These observations are associated with program \# 13003.}
\altaffiltext{1}{University of Michigan, Department of Astronomy, 1085 South University Avenue, Ann Arbor, MI 48109, USA}
\altaffiltext{2}{Observational Cosmology Lab, NASA Goddard Space Flight Center, Greenbelt MD 20771, USA}
\altaffiltext{3}{Department of Astronomy \& Astrophysics, The University of Chicago, 5640 S. Ellis Avenue, Chicago, IL 60637, USA}
\altaffiltext{4}{Kavli Institute for Cosmological Physics at the University of Chicago, USA}
\altaffiltext{5}{MIT Kavli Institute for Astrophysics and Space Research, 77 Massachusetts Ave, Cambridge, MA 02139, USA}
\altaffiltext{6}{ArmenTeKort, Antwerp, Belgium}
\altaffiltext{7}{Department of Astronomy, University of Massachusetts--Amherst, Amherst, MA 01003, USA}
\altaffiltext{8}{Department of Physics, University of Connecticut, Storrs, CT 06269, USA}
\altaffiltext{9}{Department of Astronomy, University of Texas, Austin, TX 78712, USA}
\altaffiltext{10}{Space Telescope Science Institute, 3700 San Martin Drive, Baltimore, MD 21218, USA}
\altaffiltext{$\dagger$}{Hubble Fellow}

%==================================================================================
%   ABSTRACT
%==================================================================================
\begin{abstract}
We present measurements of the surface density of star formation, 
the star-forming clump luminosity function, and the clump size distribution function, 
for the lensed galaxy \arcnamelong\ at a redshift of $z=$\zA .
The physical size scales that we probe, radii \rangeofscales, 
are considerably smaller scales  than have yet been studied at these redshifts.  
The star formation surface density we find within these small clumps is consistent with
surface densities measured previously for other lensed galaxies at similar redshift.  
Twenty-two percent of the rest-frame ultraviolet light in this lensed galaxy arises
from small clumps,  with $r<$100~pc. 
Within the range of overlap, the clump luminosity function  measured for this 
lensed galaxy is remarkably similar to those of $z\sim0$ galaxies.  
In this galaxy, star-forming regions smaller than 100~pc---physical scales not 
usually resolved  at these redshifts by current  telescopes---are important locations 
of star formation in the distant universe.  If this galaxy is representative, 
this may contradict the theoretical picture in which the critical size scale for star
formation in the distant universe is of order 1 kiloparsec.
Instead, our results suggest that current telescopes have not yet resolved the critical size scales of star-forming activity 
in galaxies over most of cosmic time.  
\end{abstract}

\keywords{galaxies: star formation --- gravitational lensing: strong --- ultraviolet: galaxies}

%==================================================================================
%   INTRODUCTION
%==================================================================================
\section{Introduction}\label{sec:intro}

Deep field surveys with \textit{The Hubble Space Telescope}  (\hst) 
have revealed that more than half of star-forming galaxies 
at $1\ga z \ga 3$ exhibit clumpy morphologies in the rest-frame 
ultraviolet \citep{Shibuya:2016}.  Motivated by these  results over 
the past decade,  a theoretical picture has emerged in which 
1~kiloparsec is a critical size scale, perhaps {\it the} critical scale, for 
star formation in the distant universe 
\citep{Elmegreen:2005fv, Elmegreen:2007id, Elmegreen:2009kd, ForsterSchreiber:2011by, 
Guo:2011cn, Guo:2015dr}.  
In that scenario, such large clumps arise from gravitational instabilities in gas-rich disks 
\citep{Toomre:1964fe, Noguchi:1999gm, Genzel:2011cp},
and are thought to highlight spots where cold gas may have accreted onto the disk 
\citep{Keres:2005gb, Dekel:2006cn, Brooks:2009bm}.

In this scenario, star formation at early times occurred in complexes that are preferentially much larger 
than in galaxies in the local universe.  
However, this hypothesis is hard to test, as structures much smaller than 1~kpc cannot 
normally be resolved by current telescopes.  
Even \hst\ is unable to  resolve star formation  at these redshifts on spatial 
scales smaller than about 500~pc, due to the diffraction limit.  
As a result,  the clumpy, 
complex morphology of star formation that is known to occur 
in nearby galaxies is normally inaccessible at the epoch when 
most of the stars in the universe were formed (see review by \citealt{Madau:2014gt}.) 

A way to overcome current observational limits and test this theoretical picture is to use
gravitational lensing by natural telescopes \citep{Einstein:1936jq, Zwicky:1937dm}.
Distant galaxies can be strongly gravitationally lensed into giant arcs that are highly magnified.  
To date, this technique has been exploited for dozens of lensed galaxies at 
$z>1$, which have revealed the importance of star-forming clumps with spatial 
scales down to several hundreds parsecs (e.g.~\citealt{Livermore:2012gw, Jones:2010hp}.)

Our  \textit{HST} program  (GO 13003) observed 
$\approx 70$ giant arcs behind 
37 lensing clusters that were selected by the 
Sloan Giant Arcs Survey (SGAS; Gladders et al., in prep).  
SGAS is a survey for strongly lensed, highly magnified galaxies selected from 
the Sloan Digital Sky Survey (SDSS).  
Galaxy clusters were selected from the SDSS photometric catalog using the 
red sequence method \citep{Gladders:2000ca}; cluster fields were 
then systematically searched for giant arcs.
Candidates lenses were confirmed, then followed up with an extensive
ground- and space-based observational campaign.
A full description of the SGAS-HST program will appear in Sharon et al.\ (2017, in prep). 
This Letter concerns one target from that larger program, \arcnamelong, hereafter \arcname.

\begin{figure*}
\includegraphics[width=6.7in,angle=0]{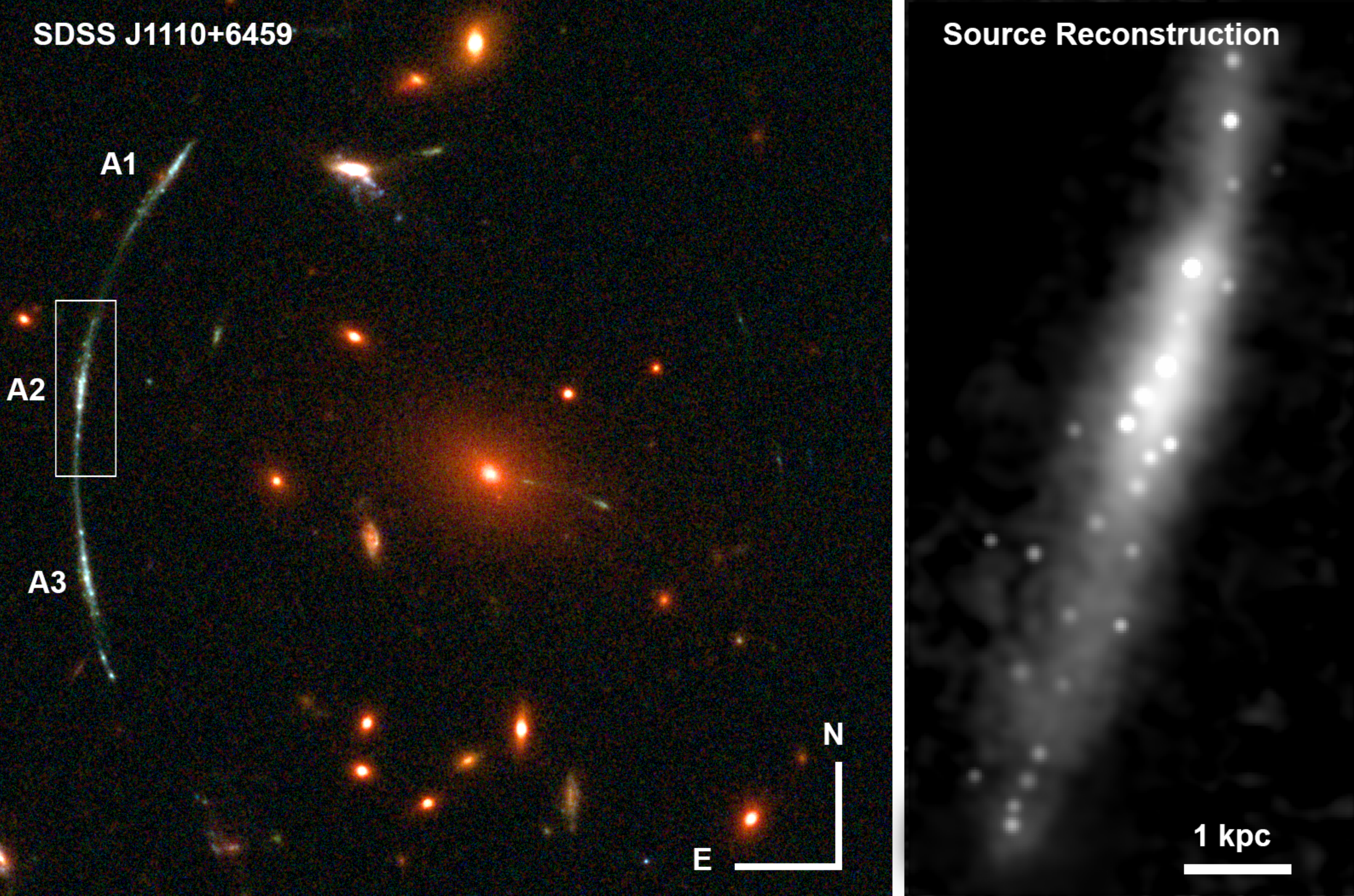} %Figs/PRfig1_try2.png}
\caption{The lensed galaxy \arcnamelong . The left panel shows
the \hst\ imaging in filters F105W, F606W, and F390W, with the three images
of \arcname\ labeled.  Image A2, the most highly magnified, is highlighted
with a box. The right panel shows our reconstruction of this lensed
galaxy in the source plane.   Two dozen clumps of star formation 
are obvious in the reconstructed image; all have sizes much smaller
than the kiloparsec scales typically probed by unlensed surveys of distant galaxies, 
and several times smaller than previously probed by gravitational lensing.
}
\label{fig:HSTimage}
\end{figure*}

In this Letter, we use the gravitationally lensed galaxy 
arcname\ to provide a 
remarkably sharp view into how stars form in the distant universe.
The giant arc, at redshift $z=2.481$, is comprised of three images of the galaxy with a 
total magnification of \totalmagnification\  (Figure~\ref{fig:HSTimage}.) 

High spatial resolution, provided by rest-frame UV imaging with \textit{HST} plus 
lensing magnification,  and combined with a new forward-modelling technique 
(Johnson et al. 2017, hereafter Paper~I),
reveals that the morphology of star formation is extremely clumpy.  
There is also a spatially extended component with rest-frame UV color indistinguishable 
from that of the clumps  (Rigby et al.\ 2017, hereafter Paper~II).
\arcname\ is forming stars at a rate of  $8.5$$^{+8}_{-0.4}$~$^{+4}_{-2}$~M$_{\odot}$ yr$^{-1}$ 
(uncertainties from SED fitting, and from the magnification uncertainty; Paper~II.)
The galaxy's stellar mass is $\log M^* = 9.24$~\Msol, with associated uncertainties of 
 $^{+0.11}_{-0.15}$  from SED fitting and 
$^{+0.08}_{-0.12}$ from the magnification uncertainty (Paper~II).
This stellar mass is comparable to the median $\log M^* = 9.4$~\Msol\ for the 
lensed sample with H$\alpha$ measurements from \citet{Livermore:2015ck}.
The peak contribution to the global star formation rate density at $z\sim2$ comes 
from galaxies with about ten times higher stellar mass, with galaxies of the stellar 
mass of \arcname\ contributing three to five times less \citep{Leja:2015fx}.
Were \arcname\ not lensed, \hst\ would measure it to 
be forming stars in a smooth, exponential disk, with a size and structure that is 
typical for galaxies of its redshift and stellar mass (Paper~II).

Paper~I produced a lens model of 
galaxy cluster \cluster, and also 
developed a novel forward--modeling technique that reconstructs 
the sizes and brightnesses of the clumps in the source plane.  This method has 
advantages over traditional ray-tracing techniques in that it is able 
to effectively deconvolve the source from the ``lensing point spread function'' (PSF),
which results as a combination of the telescope/instrumental PSF and
asymmetric shear of lensing.  
Paper~I used extensive simulations to determine the 
detectability of clumps as a function of intrinsic physical size and luminosity.  

Accordingly, with star-forming clumps detected with sizes down to \smallestscale, 
\arcname\ provides the sharpest view of a $z\sim2$ galaxy yet obtained.
In this Letter, %using the lensing model and source-plane reconstructions of Paper~I, 
we analyze the distribution of star formation surface density within \arcname, 
as well as the clump luminosity function and size distribution function.
In doing so, we probe, for the first time, 
star formation at cosmic noon on spatial scales well below 100~pc.

\section{Methods}\label{sec:methods}
This Letter builds on the results from Paper~I and Paper~II.
In Paper~I, we constructed a source-plane model of the unlensed image
of each of the emission clumps in \arcname, 
modeling each clump as a Gaussian 
parameterized by its half-width-at-half-maximum (HWHM) size 
and its intrinsic flux normalization. 
Paper~I also provides an estimate of the  
$80\%$ flux completeness limit of 
\fluxcompleteness\ (the magnitude at which artificially injected clumps were recovered 
$80\%$ of the time), and quantifies the 
smallest spatial scales that can be distinguished in the source plane
due to lensing PSF.
The constraints on the stellar populations that are used in this
Letter were derived in Paper~II,
from a spectral energy distribution analysis. 

In this work, we estimate the star formation rates (SFR) of the clumps in
\arcname\ as follows.  
To obtain the intrinsic rest-frame ultraviolet flux density, we integrate 
the 2D Gaussian that best fits the F606W source-plane reconstruction of each clump, 
and divide by ($1+z$) to correct for bandwidth compression.  
We use Equation~1 of \citet{Kennicutt:1998ki}
to estimate the SFR, 
with a correction factor of 1.8 to convert the initial mass function from Salpeter to Chabrier.
In calculating SFRs, we assumed no extinction, 
given the constraints of $A_v = 0.0$--0.2 derived from fitting the spectral 
energy distribution (Paper~II).

In the following sections, we compare our results to
measurements of SFR inferred from Paschen~$\alpha$ luminosities
of H~II regions in nearby galaxies \citep{Liu:2013fw}.
To convert the Paschen~$\alpha$ inferred rate, we use Equation 2 of
\citet{Kennicutt:1998ki}, with the same correction factor for the initial mass function.
%correction factor of 1.8 to convert the IMF to Chabrier.
We take \HaParatio\ as the intrinsic H$\alpha$/Pa$\alpha$ ratio, 
which assumes Case B recombination, $T=10^4$~K, and $n_e =100$~\cc .

We assume a flat cosmology with $\Omega_M = 0.3$, $\Omega_{\Lambda} = 0.7$, 
and $H_0 = 70$~\kms~Mpc$^{-1}$.
In this cosmology, an angular size of 1\arcsec\ 
corresponds to an angular diameter distance of 8.085~kpc at the 
redshift of \arcname\ at $z=$\zA . 

\section{Results}\label{sec:results}
The physical properties of individual star-forming clumps in
\arcname\ can be compared to measurements of other galaxies from the
literature, by examining the size--SFR space, and the
distributions of sizes and SFRs of individual clumps or galaxies. 

In Figure~\ref{fig:surfacedensity}, we consider the surface density of star formation
in the individual clumps in \arcname.
This plot, developed by \citet{Livermore:2012gw} and \citet{Livermore:2015ck}, 
is a way to parameterize the intensity of star formation.
We compare to literature measurements of galaxies at
a range of redshifts, from $z=0$ to $z=5$, both lensed and unlensed, 
mostly compiled by  \citet{Livermore:2015ck}.

For \arcname, the SFR is inferred from the rest-frame ultraviolet
(F606W filter) as explained in \S~\ref{sec:methods}.
In the comparison samples, the star formation rates for most of the
measurements come from H$\alpha$, 
with the exception of the \citet{Swinbank:2007er} and
\citet{Swinbank:2009bb} samples, 
which use the [O II]~3727, 3729~\AA\ doublet. 
Local measurements come from the SINGS galaxy sample \citep{Kennicutt:2003jt} and 
the DYNAMO $z\sim 0.1$ sample \citep{Fisher:2016jp}.  
Relative to \citet{Livermore:2015ck}, we have changed how the $z=0$ SINGS galaxies 
are plotted; since \citet{Livermore:2015ck} did not probe as small
clumps sizes as we do here, they had binned each galaxy and applied a surface brightness threshold,  
to match the spatial resolution and depth of their sample of lensed galaxies.
Here, we re-measure the SINGS galaxies without binning.  % Email from Rachael 13 July 2016
We note that of the $z\ga 1$ galaxies, the % 
samples of  \citet{Swinbank:2012gs} and %WiggleZ 
\citet{Wisnioski:2012cba} samples are not gravitationally lensed, whereas 
the other distant galaxy samples are lensed. 

\begin{figure*}
\includegraphics[width=6in]{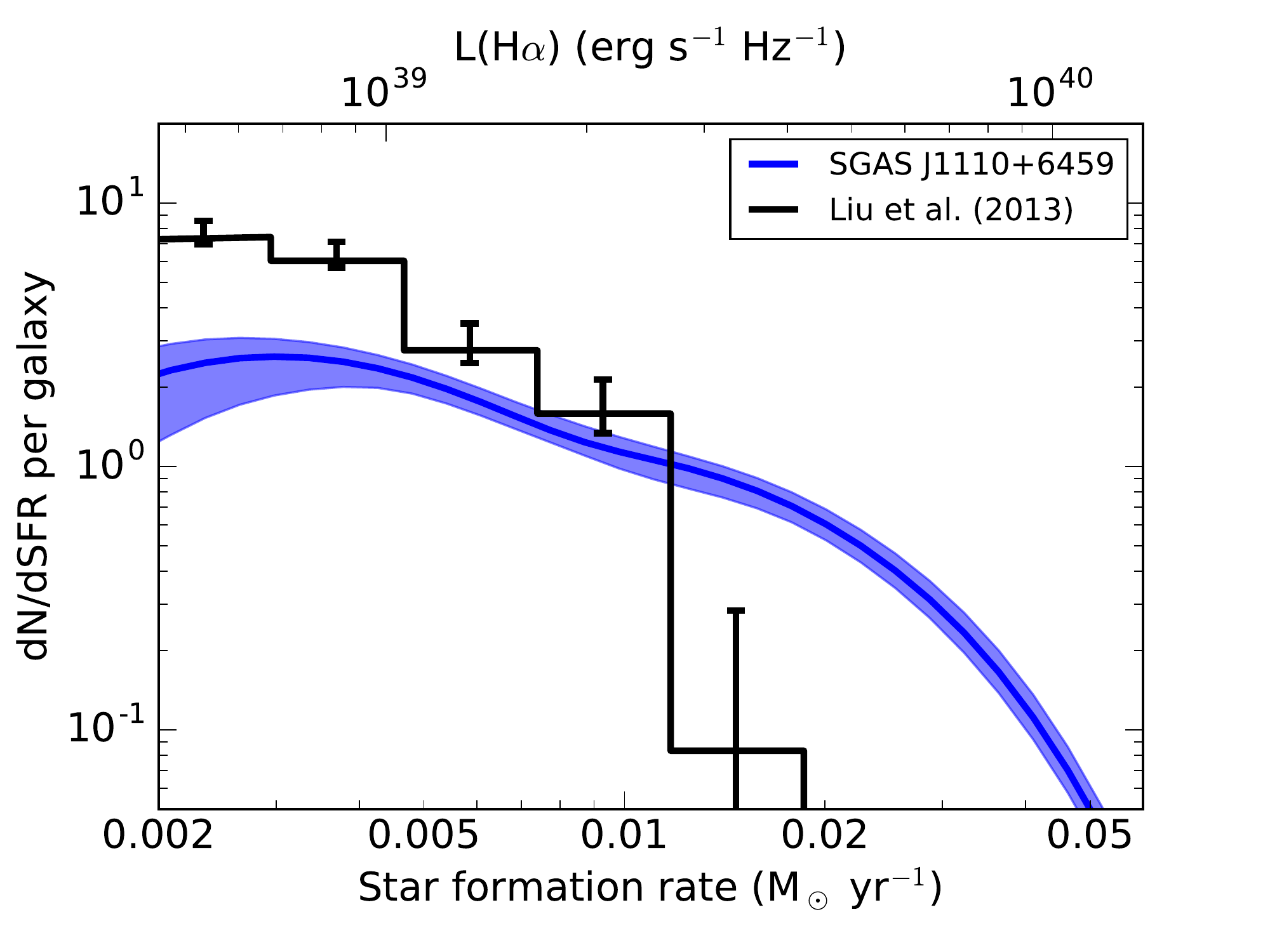} %{Figs/SFR_distribution_liu.pdf} 
\caption{The differential luminosity function of star-forming clumps.  
The median of the incompleteness-corrected 
aggregate posterior distribution function of clump
luminosity, for clumps above our 80\% completeness limit, for the 
F606W filter for  \arcname\ are plotted as the blue line;
this is the kernel density estimate for the posterior probability 
density function, corrected for incompleteness based on the 
Markov Chain Monte Carlo forward modeling described in Paper~I, 
and normalized to the number of clumps per galaxy.
The shaded region shows the 16th and 84th percentiles.
The black steps show the corresponding normalized distribution for H~II regions in the nearby 
universe \citep{Liu:2013fw}, measured using Paschen $\alpha$, %from Figures 10 and 11 
 and normalized by the number of galaxies.
}\label{fig:sfr_distrib}
\end{figure*}

\begin{figure*}
\includegraphics[width=6in]{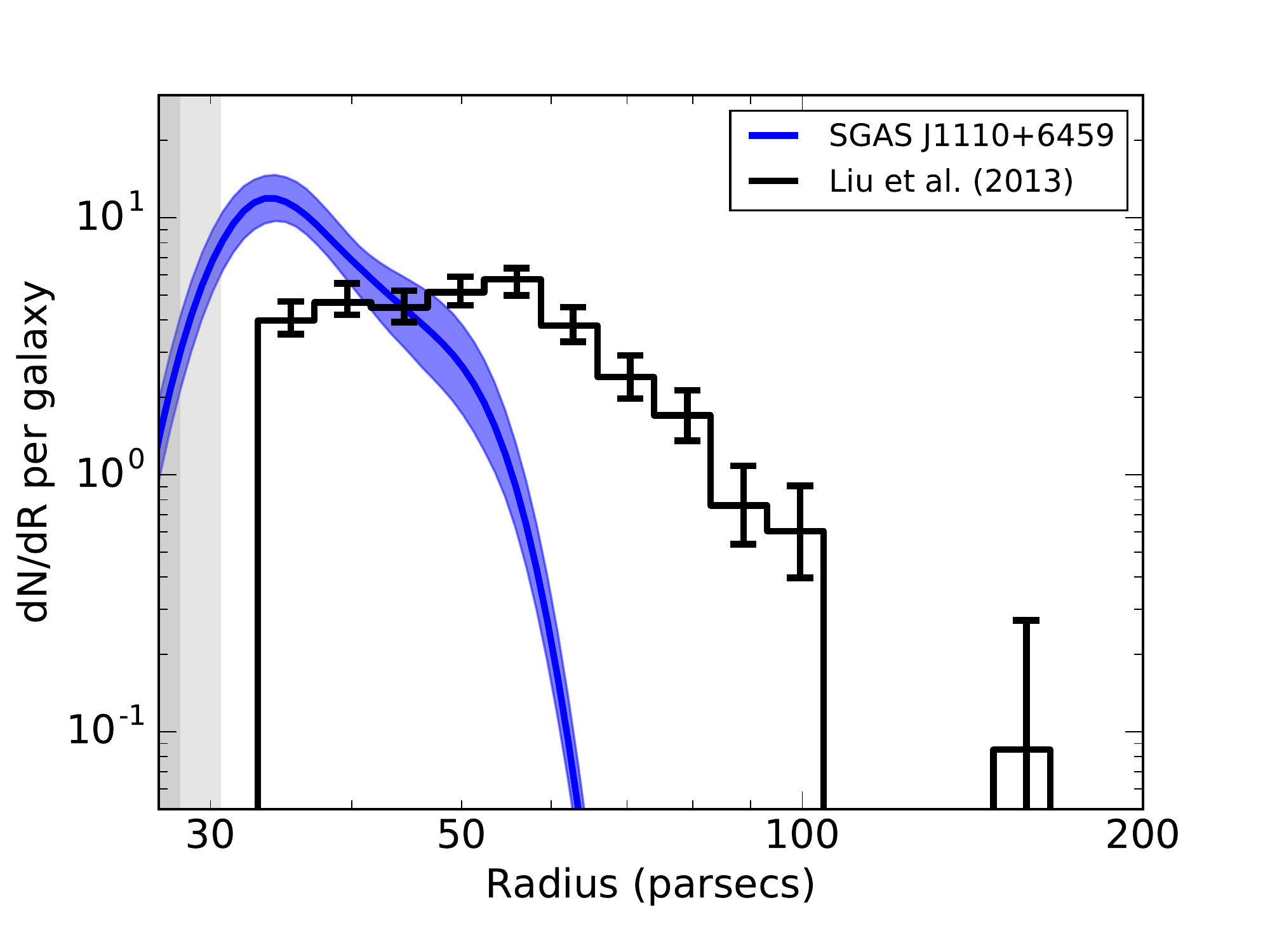} %{Figs/size_distribution_liu.pdf}
\caption{The differential distribution function of star-forming clump size (radius). 
The median of the incompleteness--corrected posterior probability 
density function for \arcname, filter F606W, is plotted as the blue line. 
The shaded region shows the 16th and 84th percentiles.
Radii for \arcname\ are half width at half maximum (HWHM).
Radii from the $z\sim0$ comparison sample \citep{Liu:2013fw} are 
isophotal  from HIIphot \citep{Thilker:2000dt}.
Given these different techniques for measuring size,
we expect a normalization offset; 
for lensed galaxies the isophotal sizes can be $\sim25\%$ larger than 
Gaussian sizes \citep{Livermore:2012gw}.
 The vertical stripes are the lensing PSF limits.  
Stochastic effects are likely to blame for the mismatch at large sizes.
}
\label{fig:size_distrib}
\end{figure*}

In Figures~\ref{fig:sfr_distrib} and~\ref{fig:size_distrib}, we plot the differential luminosity
function of clumps in \arcname, and differential distribution of clump
sizes, respectively, and compare them to measurements from the nearby universe \citep{Liu:2013fw}.

We measure the SFR surface density at $z\sim2.5$ for star-forming clumps 
with radii below 100 pc.  To the best of our knowledge, this is the first time 
that size scales well below 100~pc have been measured in the distant universe. 
In Figure~\ref{fig:surfacedensity}, the clumps from \arcname\ 
follow the $1.5<z<3$ line of constant star formation surface density 
that has been measured \citep{Livermore:2015ck},  for larger size scales, at this epoch.
Thus, we confirm previous results \citep{Livermore:2012gw, Livermore:2015ck} 
that star-forming clumps in bright lensed
$z\sim2$ galaxies have high star formation surface density, 
considerably higher than observed in the  $z\sim 0$ 
SINGS sample \citep{Kennicutt:2003jt} comparison sample, and 
comparable to those observed in the $z\sim 0$ \citet{Fisher:2016jp} sample.

Twenty-two percent of the rest-frame ultraviolet light of \arcname\ arises from 
more than twenty star-forming clumps, with measured sizes of  \rangeofscales. 
The largest clump we measure is \largestclump.  

The clump luminosity function shows a similar slope 
and normalization to that observed for Pa$\alpha$ at $z\sim0$.  
Figure~\ref{fig:sfr_distrib} shows reasonably good agreement, 
in the range of overlap, 
between the luminosity function we measure for \arcname\ and that measured
using Pa$\alpha$ for nearby galaxies by \citet{Liu:2013fw}.  A detailed comparison is
premature given that we are only examining one galaxy, but this result
indicates the value of expanding such comparisons to larger samples of lensed galaxies.

 The distribution function of clump size (Figure~\ref{fig:size_distrib}) 
is dominated by smaller clumps 
($r\sim 30$--40~pc), and shows a notable lack of the larger ($40<r<100$~pc) clumps 
that dominate the size distribution seen for P$\alpha$ at 
$z\sim0$ \citep{Liu:2013fw}.  
Simulations (Paper~I) indicate that the smallest size scale we can recover in \arcname\ is 
about \smallestscale.
Stochasticity is doubtless a factor at the larger sizes, given that we are only examining one galaxy,
and prevents any comparison of the slope with the power law observed at larger sizes
by \citet{Liu:2013fw}.
 
\section{Discussion}
\begin{figure*}
\includegraphics[width=6.5in,angle=0]{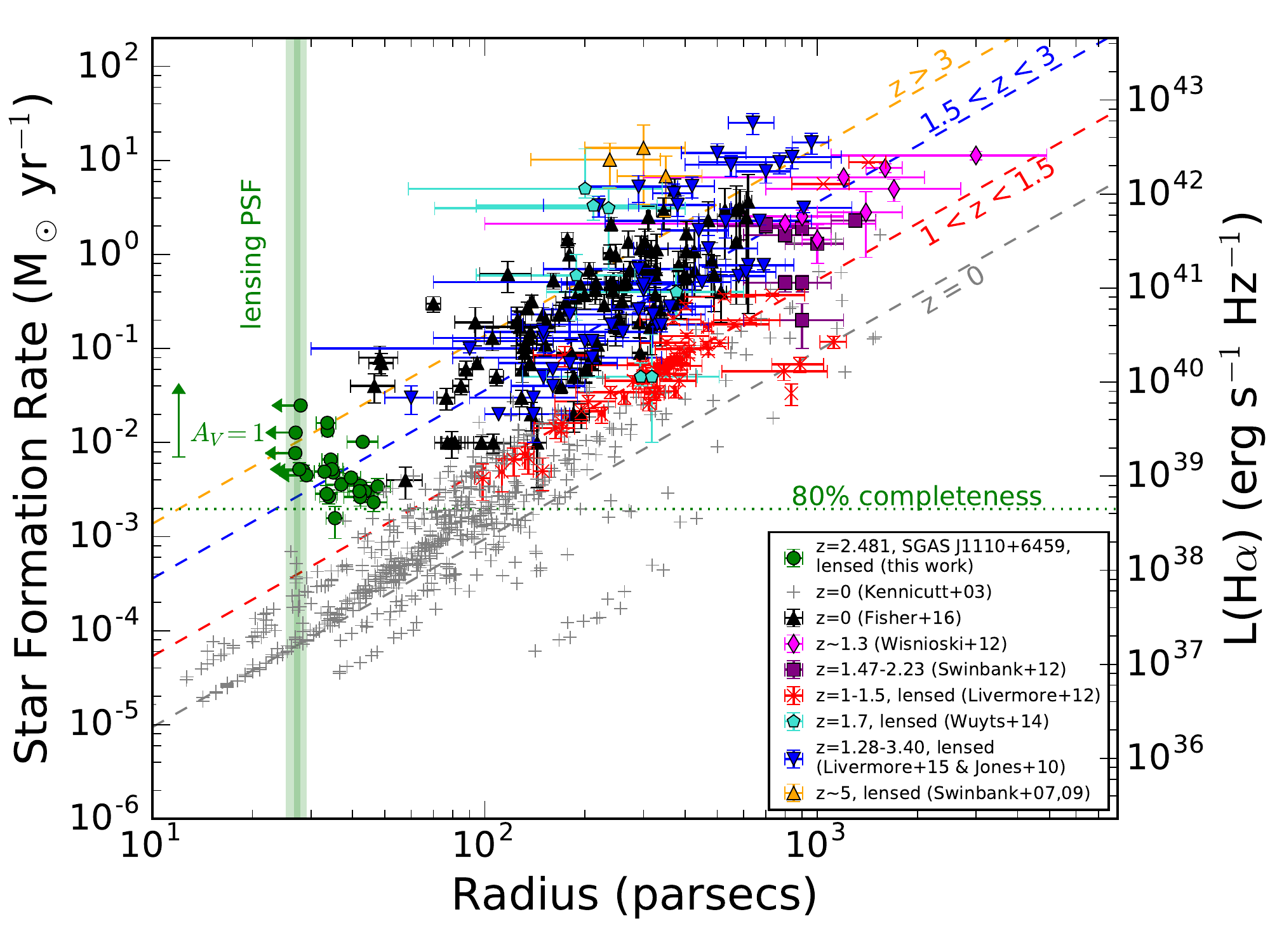} %{Figs/SFR_vs_size_606_Mar10.pdf}
\caption{The star formation -- radius relation.  
In green, we plot the results from this work:
clumps from lensed galaxy  \arcnamelong\ at z$=$\zA,  
with star formation rates and sizes estimated 
from the rest-frame ultraviolet (F606W filter).  A reddening vector shows how the 
star formation rates inferred from the rest-frame ultraviolet would increase 
due to 1 magnitude of extinction (A$_v=1$).
The vertical green stripe is the lensing PSF, with the interquartile range 
as the inner region, and the full range measured as the outer region. 
In the five cases 
where the lensing PSF was larger than the measured size,
we plot the size as an upper limit, set at the lensing PSF. 
The horizontal dashed line is the $80\%$ completeness limit determined
in Paper~I, corresponding to a source plane flux of \fluxcompleteness.
Comparison samples from the literature are over-plotted 
\citep{Swinbank:2007er, Swinbank:2009bb, Jones:2010hp, Wisnioski:2012cba, 
Livermore:2012gw, Swinbank:2012gs, Wuyts:2014eu, Livermore:2015ck, Fisher:2016jp}; 
for most,   the star formation rate was measured using H$\alpha$; 
for the $z\sim5$ galaxies [O II] 3727 was used instead.  
Diagonal dashed lines show the best fit relations in four redshift bins
from \citet{Livermore:2015ck}.
This figure is adapted from \citet{Livermore:2012gw} and \citet{Livermore:2015ck}, 
though unlike that work, we do not filter the SINGS H$\alpha$ images 
\citep{Kennicutt:2003jt} to 
match the literature measurements at $z\sim1$--1.5, but instead
use those images at their native resolution and depth.
Of the two$z\sim0$ samples, \citet{Kennicutt:2003jt}  was chosen 
to represent ``normal'' galaxies, while \citet{Fisher:2016jp} have  
higher star formation rates.
}
\label{fig:surfacedensity}
\end{figure*}

Previous measurements \citep{Livermore:2012gw, Livermore:2015ck}   
of the evolution of star formation surface density  over cosmic time
shown in Figure~\ref{fig:surfacedensity} 
have been interpreted as evidence that galaxies
become more extreme with increasing redshift, evolving toward 
higher star formation surface densities.  However, there are important selection effects at work.
The lensed galaxies plotted in Figure~\ref{fig:surfacedensity} are generally of high surface brightness.
Since lensing preserves surface brightness, and surface brightness dimming scales as $(1+z)^4$, 
we would expect only regions of high surface brightness to be observable in these galaxies \citep{Calvi:2014bm}.
Indeed,  Figure~\ref{fig:surfacedensity} shows that for \arcname, 
we could {\bf not} recover clumps with the typical surface brightnesses of 
$r<100$~pc clumps from SINGS, but we {\bf could} 
recover clumps with typical surface brightnesses of the $z\sim 0.1$ DYNAMO 
sample \citep{Fisher:2016jp}.

Given this, we cannot rule out a picture in which the star formation in \arcname\ 
is even clumpier than we can measure, and that the clumps we detect are only the 
brightest with highest surface brightness.  This would explain why 
the spatially extended (``diffuse'') component shows the same 
rest-frame UV color as the clumps (Paper~II).

The previous use of SINGS \citep{Kennicutt:2003jt} as the 
$z=0$ comparison sample may have exaggerated 
the contrast between $z\sim0$ and $z\ga1$, 
since the SINGS sample, by selection, does not include vigorously star-forming galaxies. 
By contrast, Figure~\ref{fig:surfacedensity} shows that 
the clumps from the $z\sim 0.1$ DYNAMO galaxies \citep{Fisher:2016jp} 
have star formation surface densities
an order of magnitude higher than those of the SINGS galaxies, and
indeed just as high as those seen at $z \sim 2$.

In the future, it will be possible with \textit{JWST} to make apples-to-apples 
comparisons between the most luminous clumps in nearby luminous infrared galaxies, 
and the most luminous clumps in lensed galaxies, 
for example using  H$\alpha$ integral field unit spectroscopy from NIRSpec, 
or Paschen~$\alpha$ integral field spectroscopy from MIRI.

Nevertheless, the measurements presented in this Letter indicate that at the
epoch of galaxy assembly,  star formation occurred on much smaller 
spatial scales than has been previously assumed.  
Such spatial scales have not been 
previously accessed in the distant universe, as without lensing
magnification, they fall below the resolution limit of present day telescopes.

The lensed galaxy \arcname\ at $z=$\zA\  is forming stars at a rate of 
$8.5$$^{+8}_{-0.4}$$^{+4}_{-2}$~M$_{\odot}$ yr$^{-1}$ (uncertainties from SED fitting,
and from the magnification uncertainty; Paper~II.)
High spatial resolution, provided by rest-frame UV imaging of \textit{HST} plus 
lensing magnification, reveals that about  $22\%$ of this star formation occurs
in more than twenty star-forming knots with characteristic
sizes of   \rangeofscales .
The rest of the star formation occurs in a spatially extended component 
with a rest-frame UV color indistinguishable from that of the clumps
(Paper~II).

The star formation rate surface density of the clumps in \arcname\ 
is consistent with previous measurements
at $z\sim2$ for other lensed and unlensed galaxies.  What is new is that much of the 
star formation is seen to  occur on spatial scales as small as 30~pc---spatial scales not 
previously accessed in the distant universe, as without lensing
magnification, they would be smaller than the resolution limit of
present day telescopes.

To the best of our knowledge, the only other estimates for star
forming regions as small as the ones we are measuring at these
redshifts come from observations of the Frontier Fields \citep{Lotz:2017ic}, 
using 140-orbits of \hst\ per cluster.
\citet{Vanzella:2017wd,Vanzella:2017ea} report the discovery of three young, compact star clusters
at redshift $z\sim3.2$, behind the lensing clusters MACS~J0416 and AS1063, 
with effective radii of $R_e \sim30-80$ pc.

In the local universe, star formation on such small spatial scales
would not be surprising (c.f. \citealt{Liu:2013fw},  Larson et al.\ in prep.). 
However, our result runs contrary to a theoretical picture that has emerged over the past decade: 
that 1~kiloparsec is a critical size scale, perhaps {\it the} critical scale, for 
star formation in the distant universe 
\citep{Elmegreen:2005fv, Elmegreen:2007id, Elmegreen:2009kd, ForsterSchreiber:2011by, 
Guo:2011cn, Guo:2015dr}, driven by gravitational instabilities in gas-rich disks 
\citep{Toomre:1964fe, Noguchi:1999gm, Genzel:2011cp},
in sites of cold gas accretion \citep{Keres:2005gb, Dekel:2006cn, Brooks:2009bm}.
Our measurements form a cautionary counter-example to this picture,
indicating that star formation can happen on much smaller scales in the distant universe.

This Letter examines just one of the $\sim70$ SGAS lensed galaxies 
imaged by \textit{HST}.  Future application of the techniques of Paper~I to the full sample 
will provide a much richer picture of star formation on small spatial scales at high redshift.

In conclusion, the exceptionally fortuitous lensing geometry in \arcname,
combined with new lensing methods (Paper I), 
provide an opportunity to probe spatial scales as yet
unresolved by \hst\ at this redshift.
If the nature of \arcname\ is typical of its epoch,
then much of the star-formation in the distant universe may 
take place on spatial scales much smaller than 1 kiloparsec. 
Our lensing--assisted measurements of \arcname\ 
suggest that the theoretical picture of star formation in the early universe 
requires revision,  and that most star formation in the distant universe awaits 
resolution by  future UV/optical space telescopes.

\acknowledgements
This paper is based on observations made with the 
NASA/ESA Hubble Space Telescope, obtained at the Space Telescope Science Institute, 
which is operated by the Association of Universities for Research in Astronomy, Inc., 
under NASA contract NAS 5-26555. 
These observations are associated with \hst\ program \# 13003.
Support for \hst\  program \# 13003 was provided by NASA through a grant from
the Space Telescope Science Institute, which is operated by the
Association of Universities for Research in Astronomy, Inc., under
NASA contract NAS 5-26555. 
TLJ acknowledges support by NASA under Grant Number NNX16AH48G.
KEW acknowledges support by NASA through Hubble Fellowship grant
\#HF2-51368 awarded by the Space Telescope Science Institute, which is
operated by the Association of Universities for Research in Astronomy,
Inc., for NASA, under contract NAS 5-26555.
JRR thanks the late Fred Lo for useful discussion of these results.

\bibliographystyle{astroads}
\bibliography{\string~/Dropbox/papers}
\bibliography{papers}

 \end{document}